\documentclass[twocolumn,aps,prl]{revtex4-1}
\usepackage{times}
\usepackage{graphicx}
\usepackage{amsfonts}
\usepackage{amsmath, amsthm, amssymb}
\usepackage{dsfont}
\usepackage{color}
\usepackage{empheq}
\usepackage{standalone}
\usepackage[most]{tcolorbox}
\usepackage{physics}
\usepackage{bm}
\usepackage[colorlinks=true, allcolors=black, citecolor=blue, linkcolor=blue]{hyperref}
\usepackage[utf8]{inputenc}
\usepackage[margin=0.8in]{geometry}
\usepackage{amssymb,amsmath,amsfonts,amscd}
\usepackage{mathrsfs}
\usepackage{siunitx}
\usepackage{commath}
\usepackage{graphicx}       
\graphicspath{ {Images/} }  
\usepackage{braket}
\usepackage{empheq}
\usepackage[most]{tcolorbox}
\usepackage{physics}

\definecolor{DarkRed}{rgb}{0.80,0,0}
\definecolor{lightBlue}{rgb}{0.5,0.63,0.9}
\definecolor{green}{rgb}{0,0.8,0.6}

\usepackage{bibunits}
\defaultbibliography{Refs}
\defaultbibliographystyle{apsrev4-1}

\usepackage{hyperref} 
\hypersetup{
    colorlinks=true,
    linkcolor=black,    
    citecolor=black,    
    urlcolor=blue,      
}

\usepackage{soul}

\allowdisplaybreaks[1]

\begin{document}

\title{Going beyond the Chandrasekhar-Clogston limit in a flat-band superconductor}

\author{Atousa Ghanbari}
\thanks{These authors contributed equally to this work}
\affiliation{\mbox{Center for Quantum Spintronics, Department of Physics, Norwegian University of Science and Technology,}\\NO-7491 Trondheim, Norway}

\author{Eirik Erlandsen}
\thanks{These authors contributed equally to this work}
\affiliation{\mbox{Center for Quantum Spintronics, Department of Physics, Norwegian University of Science and Technology,}\\NO-7491 Trondheim, Norway}

\author{Asle Sudb{\o}}
\affiliation{\mbox{Center for Quantum Spintronics, Department of Physics, Norwegian University of Science and Technology,}\\NO-7491 Trondheim, Norway}

\author{Jacob Linder}
\email[Corresponding author: ]{jacob.linder@ntnu.no}
\affiliation{\mbox{Center for Quantum Spintronics, Department of Physics, Norwegian University of Science and Technology,}\\NO-7491 Trondheim, Norway}



\begin{abstract}
The Chandrasekhar-Clogston limit normally places stringent conditions on the magnitude of the magnetic field that can coexist with spin-singlet superconductivity, restricting the critical induced Zeeman shift to a fraction of the superconducting gap. Here, we consider a model system where the spin-singlet Cooper pairing in a dispersive band crossing the Fermi level is boosted by an additional flat-band located away from the Fermi level. The boosting of the pairing in the dispersive band allows for nontrivial solutions to the coupled gap equations for spin-splitting fields considerably larger than the superconducting gaps at zero field. Further, the additional Cooper pairing in the flat-band, away from the Fermi level, can increase the superconducting condensation energy without affecting the paramagnetic susceptibility of the system, making the free energy favor the superconducting state. This opens up the possibility for spin-singlet superconductivity beyond the standard Chandrasekhar-Clogston limit.

\end{abstract}

\maketitle

\begin{figure}[t]
\hspace{-0.25in}
\includegraphics[width=1.0\columnwidth,trim= 0.0cm 2.6cm 1.1cm 0.0cm,clip=true]{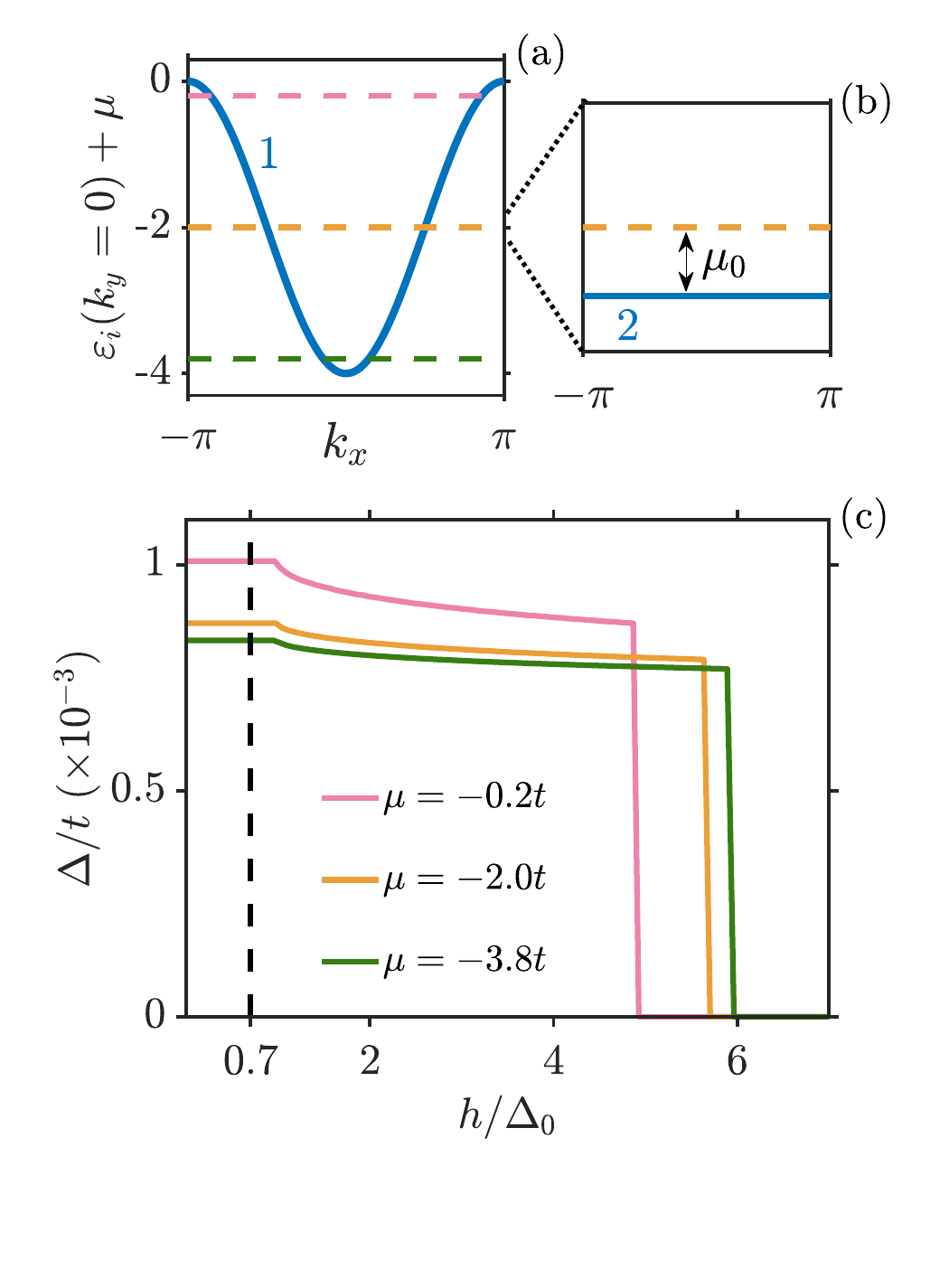}
\caption{(a-b) Illustration of the band structure of the two-band model in the absence of spin-splitting. Dashed lines represent three different values of the chemical potential $\mu = -0.2t$, $- 2t$, and $-3.8t$. The flat-band is fixed $\mu_0$ below the Fermi level, which is illustrated by the blue line 2 in (b) for a specific choice of the chemical potential. (c) Superconducting gap versus the ratio between the strength of the spin-splitting field and the gap at zero field for the three different chemical potentials in (a). The Chandrasekhar-Clogston limit is indicated by the vertical dashed line. The parameters have been set to $T = 0$, $V_{11} = V_{12} = V_{21} = V_{22} = 0.01 t$, $\mu_0 = 0.00495 t$, and $\hbar \omega_c = 0.05 t$.}
    \label{Fig1}
\end{figure}

\textit{Introduction}. -- Coexistence of superconductivity and magnetism is essential within the field of superconducting spintronics \cite{SC_spintronics_Linder_2015,SC_spintronics_Eschrig_2011, SC_spintronics_Singh_2015, SC_spintronics_Tao_2010, SC_spintronics_Buzdin_2005,SC_spin_hall_effect_Takahashi_2002, SC_spin_hall_effect_Wakamura_2015,SC_spin_hall_effect_Takahashi_2011},  which relies on stabilizing superconductors in proximity to magnetic materials and realizing phenomena such as spin-polarized supercurrents \cite{keizer_nature_2006, khaire_prl_10, robinson_science_2010}. Moreover, spin-split superconductors can give rise to very large thermoelectric effects \cite{thermoelectric_spin_splitting, SC_thermoelectric_Machon_2013, SC_thermoelectric_Giazotto_2014, SC_thermoelectric_Machon_2014, Bergeret2018, kolenda_prl_16}, which can be used to convert excess heat into useful energy. \\
\indent Magnetism is, however, usually detrimental to superconductivity. Orbital effects induced in a superconductor due to a magnetic field can be suppressed by making the superconductor sufficiently thin and applying the magnetic field in-plane \cite{Bergeret2018, Meservey_1970,Meservey_1975}. The critical magnetic field is then determined by the Zeeman-splitting that the superconducting state can survive \cite{Chandrasekhar1962, Clogston1962}. As the normal state of the system has a nonzero density of states at the Fermi level, the free energy can be lowered in the presence of a spin-splitting field by spin-polarizing the system. A spin-singlet superconductor with a gap around the Fermi level \cite{Bardeen1957}, on the other hand, has no zero-temperature paramagnetic susceptibility and is unable to lower its energy in the same way. When the Zeeman energy gain in the normal state becomes as large as the superconducting condensation energy, the system therefore transitions to the normal state. This places an upper bound on the spin-splitting field that a conventional superconductor can coexist with $h = \Delta_0/\sqrt{2} \approx 0.7 \,\Delta_0$ \cite{Clogston1962, Chandrasekhar1962}, referred to as the Chandrasekhar-Clogston limit. Here,  $\Delta_0$ is the superconducting gap at zero field. Bypassing the  Chandrasekhar-Clogston limit requires e.g.\! spin-triplet or Fulde-Ferrell-Larkin-Ovchinnikov (FFLO) pairing \cite{Fulde1964, Larkin1964}, introduction of spin-orbit coupling in the system \cite{Bruno1973}, or an applied voltage bias driving the superconductor out of equilibrium \cite{Ouassou2018}.\\
\indent Fermionic flat-band systems are systems containing one or more fermionic energy bands with weak or no dependence on momentum \cite{Leykam2018, Balents2020}. Such bands can be generated by realizing particular tight-binding models \cite{Sutherland1986, Lieb1989, Mielke_1991, Tasaki1992, Miyahara2005, Ramachandran2017, Sil2019} in e.g.\! artificial electronic lattices \cite{Tadjine2016, Qiu2016, Slot2017, Drost2017} or optical lattices filled with ultracold fermionic atoms \cite{Taie2020, Stamper2012}. For instance, spin-imbalanced superfluidity in lattices featuring flat bands, such as Lieb and kagome lattices, have been studied in \cite{Huhtinen2018, Tylutki2018}. Flat-bands can also be realized in twisted or lattice mismatched multilayers such as twisted bilayer graphene \cite{Bistritzer2011, Cao2018, Balents2020, Zhang2020, Park2021}, where the flat-bands are defined in a mini-Brillouin zone corresponding to a long-wavelength superlattice arising from the mismatch between the periodic structures in the separate layers. Flat-band systems are appealing for superconductivity as a larger density of states at the Fermi level normally leads to a larger superconducting transition temperature. Early studies identified that the presence of a flat-band could in fact give rise to a linear dependence of the transition temperature on the strength of the attractive interactions \cite{Miyahara2007, Kopnin2011}, generating hope of achieving high critical temperatures. With the discovery of superconductivity in magic-angle twisted-bilayer graphene \cite{Cao2018}, interest in flat-band superconductivity rocketed \cite{Ojajarvi2018, Holder2019, Choi2018, Lian2019, Schrodi2020}. Recently, it has also been shown that superconductivity in twisted trilayer graphene can survive in-plane magnetic fields beyond the Chandrasekhar-Clogston limit \cite{Cao2021}, which has been interpreted as an indication of spin-triplet pairing \cite{Cao2021, Qin2021}.\\
\indent In this Letter, we consider a two-band model system for a spin-split superconductor, in which a dispersive band crosses the Fermi level and a flat-band is located in the vicinity of the Fermi level. We consider both attractive intra- and interband scattering, giving rise to two coupled self-consistency equations for the spin-singlet pairing amplitudes associated with the two bands. The additional Cooper pairing in the flat-band gives rise to an increase in the condensation energy, without affecting the zero-temperature paramagnetic susceptibility of the system as long as the flat-band does not cross the Fermi level. The free energy is therefore minimized by the superconducting state beyond the Chandrasekhar-Clogston limit. Moreover, as the flat-band is located away from the Fermi level, quasiparticle excitations associated with the flat-band are energetically costly also for large spin-splitting, making the flat-band contributions to the gap equations more resilient to spin-splitting fields than the contributions from the dispersive band. We therefore find that the spin-singlet pairing in this system can survive spin-splitting fields significantly larger than the superconducting gaps at zero field. We close by discussing how the physics captured by our model can be realized in experiments.\\ 
\indent \textit{Model}. -- Our system is described by an interacting two-band Hamiltonian on the form 
\begin{align} \label{Two_band_Hamiltonian_k_1}
\begin{aligned}
    &\hspace{-0.01in}H = \!\sum_{i, \bm{k}, \sigma}\! \varepsilon_{i, \bm{k}, \sigma} c_{i, \bm{k}, \sigma}^{\dagger} c _{i, \bm{k}, \sigma}\\
       &\hspace{-0.01in}-  \frac{1}{N}\! \sum_{i, j, \bm{k}, \bm{k'}}\! V_{ij}(\bm{k}, \bm{k'})\, c_{i, \bm{k}, \uparrow}^{\dagger}\, c_{i, -\bm{k}, \downarrow}^{\dagger}\, c_{j, -\bm{k'}, \downarrow}\, c_{j, \bm{k'}, \uparrow}.
\end{aligned}
\end{align}
Here, $c_{i,\bm{k},\sigma}$ is an annihilation operator for an electron in band $i$ with momentum $\bm{k}$, and spin $\sigma$. The non-interacting part of the Hamiltonian describes the dispersive band with energies $\varepsilon_{1,\bm{k},\sigma} = - 2 t \big [ \cos(k_x) + \cos(k_y) \big ] - \mu - \sigma h$ and the flat-band with energies $\varepsilon_{2, \bm{k}, \sigma} = - \mu_0 - \sigma h$. The strength of the spin-splitting field is still $h$, the number of lattice sites is denoted by $N$, and $\mu$ is the chemical potential. Further, $\mu_0$ is the shift of the flat band away from the Fermi-level, where a positive $\mu_0$ corresponds to the flat-band being located below the Fermi level. With this parametrization, the Fermi level is moved relative to the dispersive band when $\mu$ is varied, while the separation of the flat-band and the Fermi level is fixed. The band structure in the absence of spin-splitting is illustrated in Fig.\! \ref{Fig1}\,(a-b). The Hamiltonian in Eq. (\ref{Two_band_Hamiltonian_k_1}) is similar to the one used in Ref.\! \cite{Miyahara2007}, which discussed boosting of the pairing in a dispersive band through the presence of a flat-band. However, no spin-splitting field was considered in Ref.\! \cite{Miyahara2007}. \\
\indent The interaction term in the Hamiltonian allows for attractive BCS-type intraband and interband scattering \cite{Suhl1959}. The interaction is taken to be attractive in a thin shell of width $2\hbar\omega_c$ around the Fermi level 
\begin{equation} \label{potential}                                                                                          
   V_{ij} (\bm{k}, \bm{k'}) = 
        \begin{cases}
            V_{ij} > 0, & |\varepsilon_{i, \bm{k}}|, |\varepsilon_{j, \bm{k'}}| \leq \hbar \omega_c,\\
            0, & \text{otherwise}.
    \end{cases}
\end{equation}
Here,
$\varepsilon_{i, \bm{k}}$ is defined from $\varepsilon_{i, \bm{k}, \sigma} = \varepsilon_{i, \bm{k}} - \sigma h$, and $V_{ij} $ is the band-dependent attractive interaction strength. In the following, we neglect any hybridization between the bands or other changes to the normal state band structure arising from the interaction, and investigate up to what values of $h$ the attractive interaction can give rise to superconductivity. \\
\indent Performing a standard mean-field theory, defining spin-singlet gaps $\Delta_i(\bm{k}) = \frac{1}{N} \sum_{j,\bm{k'}} V_{ij}(\bm{k},\bm{k}')\langle c_{j,-\bm{k}',\downarrow}c_{j,\bm{k}',\uparrow}\rangle$, and introducing the necessary Bogoliubov-de Gennes transformation, the coupled gap equations take the form
\begin{align} \label{Gap} 
\begin{aligned}
     &\Delta_i (\bm{k}) = \frac{1}{N} \sum_{j, \bm{k}'}  V_{ij}(\bm{k}, \bm{k}') \frac{\Delta_j(\bm{k'})}{2E_{j,\bm{k}'}} 
     \\
     & \times \frac{1}{2}\Bigg [\tanh(\frac{\beta }{2}E_{j, \bm{k}', \uparrow}) + \tanh(\frac{\beta}{2}E_{j,\bm{k}', \downarrow})\Bigg ].
\end{aligned}
\end{align}
Here, $E_{i, \bm{k}} = \sqrt{\varepsilon_{i, \bm{k}}^2 + |\Delta_i(\bm{k})|^2}$, the quasiparticle energies are  $E_{i, \bm{k}, \sigma} = E_{i, \bm{k}} - \sigma h$, and $\beta = 1/(k_B T)$ is inverse temperature. The free energy, which determines whether the superconducting state minimizes the free energy, is expressed as 

\begin{align} \label{free_energy}
    \begin{aligned}
         &F = \frac{1}{4} \sum_{i, \bm{k}, \sigma}  \frac{\Delta_i^2(\bm{k})}{E_{i, \bm{k}}} \tanh(\frac{\beta}{2} E_{i, \bm{k}, \sigma}) 
         \\
         &+ \sum_{i, \bm{k}} \big(\varepsilon_{i, \bm{k}} - E_{i, \bm{k}} \big) - \frac{1}{\beta}\sum_{i, \bm{k}, \sigma}\! \text{ln}\big(1 + e^{-\beta E_{i, \bm{k}, \sigma} }\big). 
    \end{aligned}
\end{align}
The first term in this expression is simply a generalization of the term $N\Delta^2/V$, which it reduces to for the case of a single electron band.\\
\indent \textit{Results}. -- For simplicity, we start with the case where all the interaction strengths are equal ($V_{11} = V_{12} = V_{21} = V_{22} = V$). In this case, the two coupled gap equations in Eq.\!~\eqref{Gap} reduce to a single self-consistent equation for the gap $\Delta = \Delta_1 = \Delta_2$. By numerically solving this gap equation and ensuring that the free energy in Eq.\! \eqref{free_energy} is minimized, we determine the value of the gap as a function of the strength of the spin-splitting field $h$. The results at zero-temperature are presented in Fig.\! \ref{Fig1}\,(c) for different values of the chemical potential $\mu$. As displayed in this figure, a non-zero superconducting gap can exist for spin-splitting fields significantly larger than the gap at zero field $\Delta_0$.\\
\indent In the more familiar case of a superconductor with a single dispersive band crossing the Fermi level, the superconducting gap vanishes when the field strength reaches the Chandrasekhar-Clogston limit and the normal state minimizes the free energy. In Fig.\! \ref{Fig1}(c), this limit is indicated by a vertical dashed line. The mechanism for this transition is easily seen from the expression for the free energy in Eq.\! \eqref{free_energy} if we limit ourselves to the contributions from $i=1$, corresponding to the dispersive band. For the superconductor, as long as the spin-splitting is smaller than the gap, all the quasiparticle energies are positive and the last term in the free energy vanishes at zero temperature. For the normal state, on the other hand, there is no gap in the excitation spectrum and the energies $E_{1,\bm{k},\sigma} = |\varepsilon_{1,\bm{k}}| - \sigma h$ can turn negative, giving rise to negative contributions from the last term in the free energy. This corresponds to a lowering of the normal state free energy through the system becoming spin-polarized. Comparing the rest of the free energy for the two phases gives rise to the condensation energy, favoring the superconducting state. When the strength of the spin-splitting field is increased, the lowering of the free energy of the normal state eventually dominates over the condensation energy, and the normal state prevails.\\
\indent In the present case, there are  additional contributions to the free energy arising from the flat band. As long as the quasiparticle energies $E_{2, \bm{k},\sigma}$ are shifted away from the Fermi level by $|\mu_0| > h$, these energies will always be positive even without a gap. At zero temperature there are then no contributions from the last term in the free energy arising from the flat-band, regardless of whether the system is in the superconducting or normal state. The effect of the flat-band on the free energy is then simply to significantly increase the condensation energy due to its large density of states. We therefore find that having a nonzero gap minimizes the free energy also beyond the Chandrasekhar-Clogston limit. Moreover, considering the dispersive band, when the spin-splitting becomes larger than $\Delta_0$, the gaps in the separate spin-bands no longer overlap and the superconducting state is able to lower its free energy by spin-polarizing the quasiparticles as discussed in Ref.\! \cite{Sarma1963}. Such "gapless" superconductivity arises from time-reversal symmetry breaking \cite{Bennemann2008, Guoya2002} and has been encountered in e.g.\! systems with magnetic impurities \cite{Abrikosov1960, Woolf1964} and in the presence of a magnetic field \cite{Maki1964, deGennes1964}. For a model with two bands crossing the Fermi level, the state where the spin-splitting is larger than the superconducting order parameter of both bands was discussed, but not found to be stable, in Ref.\! \cite{He2009}.\\ 
\begin{figure}[t]
\hspace{-0.25in}
\includegraphics[width=1.0\columnwidth,trim= 0.2cm 0.0cm 0.0cm 0.0cm,clip=true]{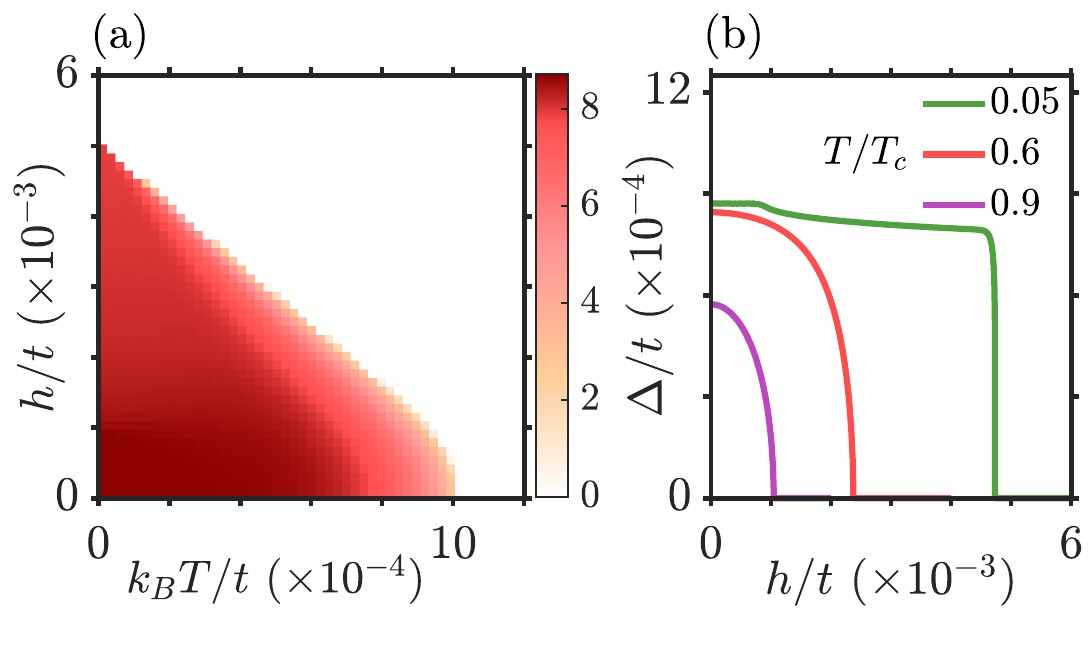}
\caption{(a) Superconducting gap $\Delta$ as a function of temperature $T$ and the strength of the spin-splitting field $h$ for the parameters in Fig.\!~\ref{Fig1} with $\mu = -2t$. (b) Curves showing $\Delta(h)$ for a selection of temperatures.}
    \label{phase_plot}
\end{figure}
\indent Turning to the gap equation, for a spin-splitting field larger than the gap, the energies $E_{1,\bm{k'},\uparrow}$ and $E_{1,\bm{k'},\downarrow}$ on the right-hand-side of Eq.\! \eqref{Gap} can end up with opposite signs, leading to a cancellation of the contributions. The first contributions to go are those with the smallest energies $E_{1,\bm{k}'}$, i.e.\! the most important contributions from the dispersive band. For the flat-band, on the other hand, the quasiparticle energies are always positive for $h < \sqrt{\mu^2_0 + \Delta^2}$. The flat-band contributions to the gap equation are therefore robust towards spin-splitting. By having the flat-band sufficiently close to the Fermi level ($|\mu_0| < V/2$), nontrivial solutions to the gap equation can then be guaranteed as long as the field is not large enough to change the sign of quasiparticle energies.\\
\indent Closer investigation of the free energy reveals that, when contributions from the dispersive band are neglected, the superconducting state is no longer favored for $h > \sqrt{\mu^2_0 + \Delta^2} - \frac{1}{2}\Delta^2/\sqrt{\mu^2_0 + \Delta^2}$. This expression is larger than or equal to $\abs{\mu_0}$ and arises from the paramagnetic energy gain of the normal state compensating the energy gain associated with the superconducting gap. Moreover, the expression is smaller than $\sqrt{\mu^2_0 + \Delta^2}$, meaning that there at this field strength still exists a nontrivial solution to the gap equation if $|\mu_0| < V/2$. The critical field is then limited by the free energy, giving rise to a first-order transition where the gap suddenly vanishes. Further, for $\mu^2_0 \gg \Delta^2$, the critical spin-splitting field simply becomes $h_c \approx |\mu_0|$, where the maximum value of $|\mu_0|$ that can produce a nontrivial solution to the gap equation is limited by the interaction strength $V$. \color{black}\\   
\indent The dependence of the gap equation on the strength of the spin-splitting field can be observed in Fig.\! \ref{Fig1}\,(c), and is most easily seen by considering the pink curve corresponding to $\mu = -0.2 t$. For $h < \Delta_0$, the curve is flat as the spin-splitting has no effect on the contributions to the gap equation. Then, as $h > \Delta_0$, contributions from the dispersive band start cancelling out, leading to a decrease in the gap. This corresponds to the minimum energy of breaking a Cooper pair becoming zero, as discussed by Abrikosov in the context of gapless superconductivity in the presence of magnetic impurities \cite{Abrikosov1988}. In the present case, a non-zero superconducting gap exists until around $h > |\mu_0|$, beyond which the free energy favors the normal state.\\
\indent Taking $\mu = - 2t$ as an example, $h_c/\Delta_0 = 5.7$ and, calculating the critical temperature at zero field, $\Delta_0/T_c = 0.87$. For these parameters, we then obtain $h_c /\mu_B = 7.4 \,\textrm{T}/\textrm{K} \times T_c$, where $\mu_B$ is the Bohr magneton. For $\mu = -0.2t$, the ratio $\Delta_0/T_c$ becomes larger as the dispersive band contributes more to the gap equation, and oppositely for $\mu = -3.8t$. Reducing $|\mu_0|$ can give rise to higher values for $\Delta_0/T_c$. Further, the temperature dependence of the results for $\mu = -2t$ are presented in Fig.\! \ref{phase_plot}. As displayed in Fig.\! \ref{phase_plot} (b), the superconductor to normal state transition becomes a second-order transition at higher temperature. The change from a first-order to a second-order transition is found to take place slightly above $T/T_c = 0.06$. This ratio can be increased by moving the flat-band closer to the Fermi level.\\
\begin{figure}[t]
\includegraphics[width=1\columnwidth,trim= 0.2cm 0.0cm 0.53cm 0.0cm,clip=true]{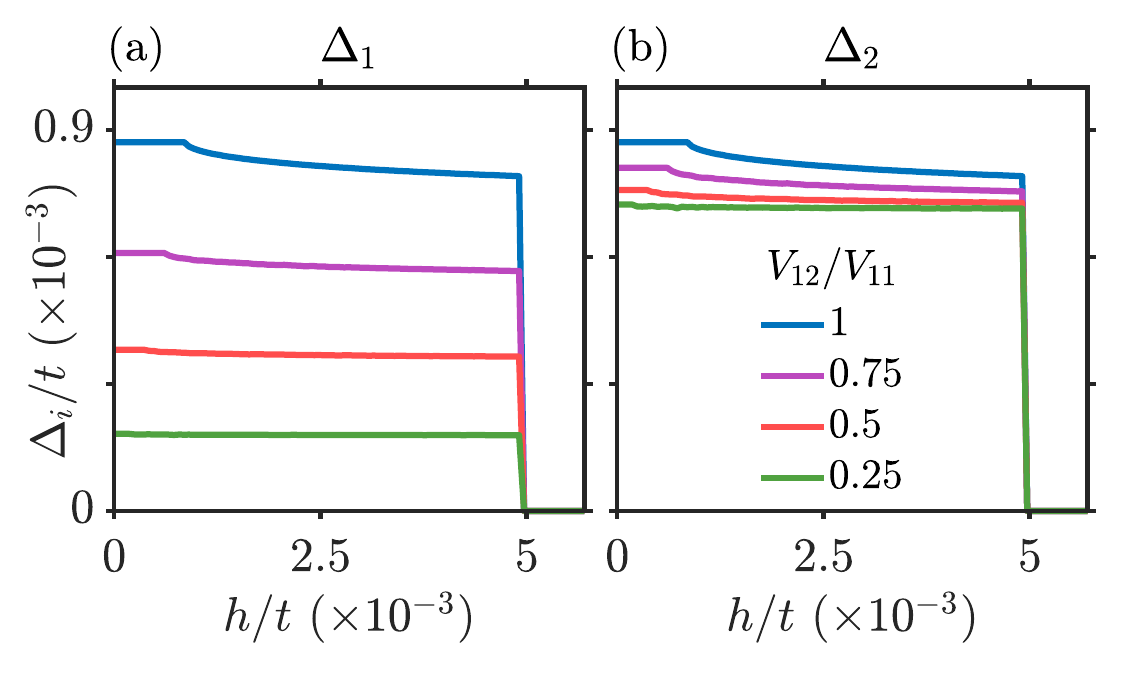}
\caption{(a) $\Delta_1$ (b) $\Delta_2$ as a function of the strength of the spin-splitting field $h$ for four different ratios of $V_{12}/V_{11}$. The parameters are set to $T = 0$, $\mu_0 = 0.00495 t$, $V_{11} = V_{22} = 0.01 t$, $V_{12} = V_{21},$ $\hbar \omega_c = 0.05 t$, and $\mu = - 2 t$.}
    \label{Fig2}
\end{figure}
\indent We next demonstrate how our results are influenced by band-dependence of the interaction strengths. We first consider the effect of reducing the interband scattering by taking $V_{12} = V_{21}$ smaller than $V_{11} = V_{22}$. Solving the coupled gap equations and checking the free energy, we obtain the results in Fig.\! \ref{Fig2}. As the dominant contributions to the gap equations arise from the flat-band, we find that $\Delta_2$, which obtains contributions from $V_{21} \Delta_1$ and $V_{22} \Delta_2$, is not strongly affected by a reduction of $V_{21}$. On the other hand, $\Delta_1$ obtains contributions from $V_{11} \Delta_1$ and $V_{12} \Delta_2$, and a reduction of $V_{12}$ therefore leads to a significant reduction of $\Delta_1$. Substantial pairing in the dispersive band crossing the Fermi level therefore requires a sufficiently large interband interaction strength. In all cases, the gaps survive until around $h > |\mu_0|$, which is considerably larger than the gaps at zero field.\\
\indent Finally, we consider the case where we also increase the intraband interaction in the dispersive band compared to the intraband interaction in the flat-band. The results for $\Delta_1$ are displayed in Fig.\! \ref{Fig3}, showing that significantly increasing $V_{11}$ only leads to a moderate increase in $\Delta_1$ as the dominant contributions to the gap equations still arise from the flat-band due to its large density of states. A moderate increase in $\Delta_1$ has little impact on the results for $\Delta_2$ which therefore varies little when we increase $V_{11}$. The gaps once again survive until around $h > |\mu_0|$, where the magnitude of $|\mu_0|$ that can still provide a nontrivial solution to the gap equations is determined by how large we take $V_{22}$.\\ 
\begin{figure}[t!]
\includegraphics[width=1.0\columnwidth,trim= 0.0cm 0.5cm 0.8cm 0.0cm,clip=true]{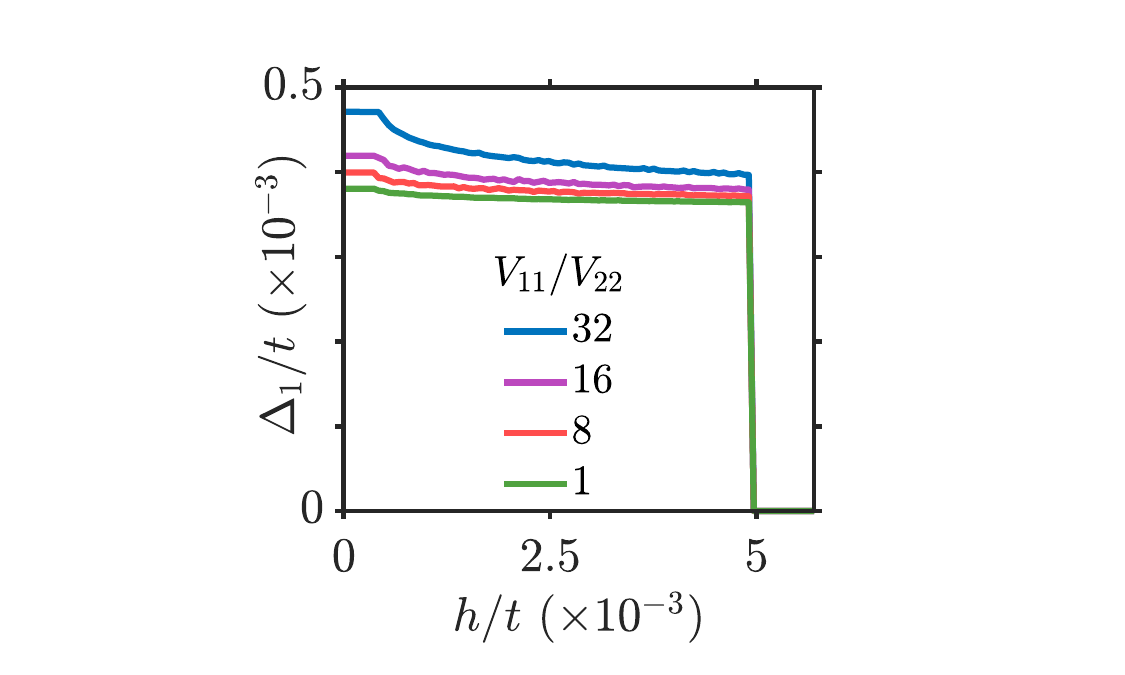}
\caption{The gap $\Delta_1$ as a function of the strength of the spin-splitting field $h$ for four different ratios of $V_{11}/V_{22}$. The parameters are set to $T = 0$, $\mu_0 = 0.00495 t$, $V_{22} = 0.01 t$, $V_{12} = V_{21} = 0.005t,$ $\hbar \omega_c = 0.05 t$, and $\mu = - 2 t$.}
    \label{Fig3}
\end{figure}
\textit{Outlook}. -- We have presented a mechanism for how a spin-singlet superconductor can survive beyond the Chandrasekhar-Clogston limit. The mechanism relies on having a sufficiently dispersive band crossing the Fermi level, an additional flat-band nearby, sufficient intra-band interaction in the flat-band, and some interband scattering. Experimental realization would typically be through a thin-film superconductor with a, preferably tunable, induced spin-splitting. The spin-splitting can be achieved by exposing the superconductor to a strong in-plane magnetic field, or to a combination of a ferromagnet and an external field where the additional external field provides the tunability of the strength of the spin-splitting \cite{Ouassou2018}. The necessary band structure could be realized in twisted multilayers, artificial electronic lattices or alternatively in optical lattices. The especially relevant case of a dispersive band on top of a flat-band corresponds to the limiting case where the chemical potential in Fig.\! \ref{Fig1} is taken almost down to the bottom of the band, e.g.\! $\mu = -4t + \mu_0$. Importantly, the flatness of the flat-band should be stable in the presence of spin-splitting. 
Finally, the interactions could originate with phonons in twisted multilayers or be engineered in artificial systems. The choice of interactions in Fig.\! \ref{Fig1} could e.g.\! in principle correspond to the electrons in both bands coupling similarly to Einstein phonons. As shown in Fig.\! \ref{Fig2} and \ref{Fig3}, the results for the critical field are, however, quite robust to band-dependence of the interaction strengths, allowing for reduction of the interband scattering as well as for a much larger intraband scattering in the dispersive band than in the flat-band.\\
\indent More exhaustive studies of realistic systems with similar properties as our model system, taking into account the details of the band structure and the interactions, should be performed in order to more closely relate the results to experiments. Special attention should be paid to the theoretical approach when a flat-band is present and when the Fermi energy is not dominating the other energy scales in the system, which e.g.\! can be the case when the chemical potential is close to the bottom of the conduction band. Future work could also include analysis of the stability of other superconducting phases such as FFLO states, or investigation of e.g.\! single-band models featuring bands that are partially flat and partially dispersive \cite{Aoki2020}, where our mechanism in principle also could be applicable.\\
\indent \textit{Summary}. -- Our results demonstrate that spin-singlet superconductivity beyond the Chandrasekhar-Clogston limit could be possible in flat-band systems. Future studies should perform more detailed calculations for realistic systems in order to more closely connect the findings to experiments.\\
\indent We thank Even Thingstad for valuable discussions. We acknowledge financial support from the Research Council of Norway Grant No. 262633 ``Center of Excellence on Quantum Spintronics'' and Grant No. 323766. \color{black}


\bibliography{Refs}
\end{document}